\newcommand{\Deg}{{}^{\,\mbox{\scriptsize o}}}		% -  degree
\newcommand{\Exp}[1]{{\mbox{\Large \rm e}}^{\normalsize \it #1}}% - exponential
\def\chem{\everymath={fam0 }\fam0 }

\documentstyle[12pt,psfig,aaspp4]{article}
\begin{document}

\title{Study of Broad Scale Anisotropy of Cosmic Ray \\
 	Arrival Directions from $2\times 10^{17} eV$ \\ 
	to $10^{20} eV$ from Fly's Eye Data}

\author{D.J. Bird\altaffilmark{1}}
\affil{Department of Physics, University of Illinois at 
	Urbana-Champaign, Urbana, IL 61801, USA }

\and
\author{H.Y. Dai\altaffilmark{2}, B.R. Dawson\altaffilmark{3},
J.W. Elbert, 
M.A. Huang\altaffilmark{4}, 
D.B. Kieda, S. Ko, E.C. Loh, M. Luo\altaffilmark{5}, 
J.D. Smith, P. Sokolsky, P. Sommers, S.B. Thomas}
\affil{Department of Physics, University of Utah, Salt Lake City, 
	UT 84112, USA }

\altaffiltext{1}{present address:Defence Science and Technology Organisation, 
	P.O Box 1500, Salisbury, S.A. 5108, Australia, 
	Australia }
\altaffiltext{2}{present address:Rosetta Inpharmatics, 12040 115th Ave NE, 
	Kirkland, WA 98034, USA}
\altaffiltext{3}{present address:Department of Physics and Mathematical Physics, University of Adelaide, Adelaide, South Australia 5005, Australia}
\altaffiltext{4}{present address:Institute of Physics, Academia Sinica, 
	11529 Nankong, Taiwan, ROC }
\altaffiltext{5}{present address:Biology Division, Dugway Proving Ground, UT 84002, USA}

\begin{abstract}
We report results on the broad scale anisotropy of cosmic ray arrival
directions in the energy rage from $2 \times 10^{17} eV$ to $10^{20} eV$. The
data was taken by the Fly's Eye detector in both monocular and stereo modes of
operation. We look for dependence on galactic latitude or supergalactic
latitude by fitting the data to a Wdowczyk and Wolfendale plane enhancement
function and a N-S gradient functional form. We report a small but
statistically significant galactic plane enhancement in the energy range
between $2\times 10^{17}eV$ and $3.2 \times 10^{18} eV$. The probability that
this anisotropy is due to fluctuations of an isotropic distribution is less
than 0.06\%. The most significant galactic plane enhancement factor $f_{E} =
0.104 \pm 0.036$ is in the energy range $0.4-1.0 \times 10^{18} eV$. No
statistically significant evidence for a N-S gradient is found. There is no
sign of significant deviation from isotropic background when the data is
analyzed in terms of supergalactic latitude distributions.
\end{abstract}

\section{Introduction}

There have been a number of reports (\cite{Gil93,Stn95}) of small 
anisotropies in the ultra high energy cosmic ray arrival direction 
distribution with respect to the galactic or supergalactic plane. Recently, a 
high statistics search for anisotropy has been reported by the Akeno/AGASA 
collaboration (\cite{Agasa98}). This group reports a small anisotropy towards 
the galactic plane at energies near $10^{18} eV$.  The statistical 
significance of this evidence is not strong, however, and the various ground 
array experiments have reported systematic errors due to temperature and 
pressure variations and detector energy scale estimation which are quite 
different from the systematic errors in Fly's Eye type of experiments. It is 
therefore important, both from the point of view of statistics and systematics,
 to have as complete a picture as possible of all the evidence on anisotropy. 
This paper is the final result on cosmic ray anisotropy from the original 
Fly's Eye experiment (\cite{Bal85,Cas85}) which ran from 1981 to 1992. 
Previous publications (\cite{paul93ani,MAH93ani}) were based on subsets of 
this complete data sample.

Observation of cosmic ray anisotropy, when taken together with studies of the
cosmic ray spectral shape and cosmic ray composition can yield very important
clues to the nature of the highest energy cosmic rays. If cosmic rays come
from the galactic plane and are largely protonic in composition, we expect to
see a galactic plane enhancement which becomes more significant with energy,
since the proton rigidity will increase. A heavy composition dominated by Iron
nuclei will tend to show much smaller anisotropy at any given energy because
of the smaller Larmor radius of the heavy nuclei. Previously reported Fly's
Eye data on spectrum (\cite{APJ94spe}) and composition
(\cite{Bird93comp,PRLcomp}) support a two component model where a mostly heavy
galactic composition is superceded by a mostly light, extragalactic
component. The cross-over between these two cosmic ray fluxes appears to be
near $3 \times 10^{18} eV$. If correct, this model would imply at best a small
galactic plane enhancement below $3\times 10^{18}eV$.

If the higher energy cosmic ray flux is indeed mostly extragalactic, it may
have a different anisotropy from the lower energy component. Strong nearby
extragalactic sources could lead to strong anisotropies at energies near
$10^{20}eV$. If extragalactic magnetic fields are not greater than a nanogauss
(\cite{ExGalB}), charged particle astronomy becomes possible at these
energies. It has been recently suggested that extragalactic sources may be
distributed along the supergalactic plane (\cite{Stn95}). If so, we might
expect an enhancement of cosmic ray arrival directions towards this plane at
high enough energies. Of course, extragalactic sources might have a more
complex distribution. They may even reflect the pattern of filaments and voids
that are seen in deep space galactic surveys (\cite{Waxman}). However, if the
origin of the highest energy cosmic rays is due to decay of topological
defects (\cite{Bhatt92,Bhatt95}), then the anisotropy may have no relation to
the distribution of visible matter in the universe.

\section{The Fly's Eye Experiment}

The Fly's eye experiment consisted of two detectors ( F.E. I and F.E. II )
located at Dugway Proving Ground ($40\Deg$ North) in Utah. The two detectors
were spaced 3.4 km apart. Details of the experimental technique can be found
in (\cite{Bal85,Cas85,MAH96}). Briefly, UHE cosmic rays entering 
the atmosphere form an extensive air shower (EAS). The ionizing particles in 
this shower excite Nitrogen fluorescence in the atmosphere. Isotropically 
emitted light from this fluorescence is collected by spherical mirrors and 
detected by photomultipliers located at the two detector sites. The 
photomultiplier tube signals can be used to determine the geometry of the EAS 
( i.e. its distance from the detector and arrival direction) and the energy of 
the primary particle. F.E. I had a larger aperture than F.E. II and events 
seen by it fall into the monocular data set. Events seen simultaneously by 
F.E.II ( which had a more limited aperture) were classified as stereo events. 
About 1/3 of the monocular data set was also seen in stereo. Because events 
seen in stereo have redundant measurements, the typical stereo geometrical and 
energy resolution is significantly better than that which is purely monocular. 
In order to have acceptable resolution, monocular data must be subjected to a 
number of cuts. The present data sample is large enought so that there is still
statistical power even after the relatively stringent cuts required.

\section{Data selection and angular resolution}
	All the Fly's Eye data used in this analysis must pass a minimum 
standard cut which cuts out events with variables outside their natural range, 
or that have error greater than one half of their range or have relative error 
greater than 10.0 (\cite{MAH96}).

	In this study, we try to minimize the angular resolution while 
keeping the number of events as large as possible. We are guided by a 
Monte-Carlo simulation using a two source model of proton showers and iron 
showers. This model gives results consistent with the Fly's Eye spectrum and 
composition results previously reported (\cite{APJ94spe,Bird93comp,PRLcomp}).
These simulated events then pass the same event reconstruction 
programs as the real data. The reconstructed value and the input value for 
different variables are then compared. The angular error is defined as the 
space angle between the input shower direction and the reconstructed one. The 
tighter the data cut used, the fewer events remain. The optimization is to 
maximize the ratio of the relative decrease of angular error and the relative 
decrease of the number of events.

	The most important factor controlling the angular error of monocular 
events is the shower track length. An event with longer track length has more 
degrees of freedom to determine arrival direction and, therefore, it has less 
angular error. We find the optimal track length is $\geq 50\Deg$. 

	The data are separated into six energy intervals E1 : $0.2-0.4EeV$, 
E2: $0.4-1.0EeV$, E3 : $1.0-3.2EeV$, E4 : $3.2-10.0EeV$, E5 : $>10.0EeV$, 
and E6 : $>32EeV$. According to the two source model, E1 should correspond to 
an almost pure iron composition while E5 and E6 consist of almost pure protons.
 The width of each energy interval is about half a decade, so we choose the 
relative energy error $\leq 3.0$. 

	A bracket of depth at shower maximum, $X_{max}$, is also applied
\[
      300 + 80.\times  log_{10}(E) < X_{max} < 1100 + 80.\times log_{10}(E)
\]
where $E$ is the energy in $EeV$. Events outside this bracket have poorly
 reconstructed shower profile and large angular errors. The relative error of 
$X_{max}$ is set at 1.0. We also apply a zenith angle cut at $\leq 80\Deg$. 

	The Fly's Eye detectors operate on moonless nights. A weather code 
(\cite{MAH96}) was recorded for every hour of observation. We required data to 
have been taken when less than 1/4 of the sky is cloudy.

	Since we don't know the actual arrival direction of real events, we
cannot define the angular error as in Monte-Carlo simulated data. We use the
uncertainty of the reconstructed value, i.e. the uncertainty in zenith angle
$d\theta$ and error in azimuth angle $d\phi$. An overall angular uncertainty
is defined as
\[	\delta= \sqrt {d\theta^2 + (\sin\theta\times d\phi)^2 } \]
This value also shows a strong dependence on track length. Statistically, the 
mean angular  uncertainty is related to the angular error 
(\cite{MAH96}). But it is impossible to predict angular error from angular 
uncertainty on an event by event basis. A cut at $15\Deg$ on angular 
uncertainty is applied to the data.

	The tight cuts we used in this study are listed in table 1. 
The resulting number of events of mono and stereo data are listed in 
table 2. Because the stereo data have only two events at energy 
$> 32EeV$, this analysis does not apply to this energy bin.

\placetable{tab1}
\placetable{tab2}

	From Monte-Carlo studies, the angular resolution is $3.2\Deg$ for 
monocular data and is $1.1\Deg$ for stereo data at the 50\% confidence level. 
At the 90\% confidence level, the angular resolution is $9.6\Deg$ for 
monocular data and is $3.2\Deg$ for stereo data.

	For the stereo subsample, we can compare the same event reconstructed 
by monocular and stereo methods. The space angle between these two direction 
is also a measure of the possible angular resolution of this subset of data. 
Figure 1 shows the resulting Poisson like distribution 
of angular resolution. At 50\% confidence level, the angular resolution is 
approximately $5.\Deg$. In the large scale anisotropy analysis, in order to have 
enough statistics in each bin,  we choose a bin width of 10 degrees for both 
monocular and stereo data. 

\placefigure{fig:ang_res_match}

\section{Isotropic background prediction}
	The background expected from an isotropic intensity can, in principle, be calculated by
\[	B(l,b,E) = \sum_{operation\; nights} \int_{T_{on}}^{T_{off}}
			R(T) \times  A( \theta, \phi, E)dT	\]

%%		     \-    /Toff
%%	bkg(l,b,E) =  >   /     R(T) \times  A(theta,phi,E) dT
%%		     /-  /Ton

where $B(l,b,E)$ is the predicted isotropic background at longitude $l$ and
latitude $b$ at energy $E$ and $R(T)$ is the event rate at time $T$. Note that
$\theta$ and $\phi$ are here regarded as functions of $l,b$ and
$T$. $A(\theta,
\phi, E)$ is the acceptance of the detector, the geometric efficiency which is 
the relative ability to detect events of energy $E$ from a certain zenith 
angle $\theta$ and azimuth angle $\phi$ . 

	Although the Fly's Eye PMT singles rate is kept constant, the real
event rate fluctuates due to variations in night sky noise and weather
condition.  Instead of using this integration, we use a scrambled event method
to determine the isotropic background with better accuracy.

	From the real data, we select the events that pass the weather code
 cut and store their trigger time in a time data bank. For those events that
 pass the tight cuts, we store their arrival direction $(\theta, \phi)$ in a
 direction data bank. The time data bank contains information on the system
 on/off time and trigger rate $R(T)$. The direction data bank contains the
 acceptance information. Then a simulated event is generated by randomly
 sampling a set of $(\theta,\phi)$ from direction data bank and an event
 trigger time from time data bank. This randomization destroys the correlation
 from any source. The simulated events thus represent an expected data set if
 cosmic rays are isotropic.

	Because of changes to the Fly's Eye hardware, the acceptance may be
different for different operation epochs. The background is calculated using
the time and direction data from the same epoch. However, at energy $>3.2
EeV$, the number of events is so small that the acceptance becomes
indistinguishable between epochs. In order to have proper statistics, we have
to combine all epochs to form a direction data bank for this energy range.

	In this study, we simulate 5,000 sets of data each having the same 
number of events as real data. The mean value of those 5,000 sets is used as 
the expected detector exposure. The fluctuation of those 5,000 sets 
represents the unceratinty in the exposure. 

\section{Broad scale anisotropy analysis}
	We compare the arrival direction distributions for two zones of 
the sky. The first zone is the whole sky while the second zone is the half of 
the sky ($30\Deg < l < 210\Deg)$ where the Fly's Eye acceptance covers most of 
the galactic latitude. We also  compare the distribution in supergalactic 
coordinates.

	The null hypothesis of this study is that cosmic rays are
isotropically distributed. Based on this hypothesis, we simulate the expected
sky distribution and then compare it with the real data distribution. The data
and background are binned in $10\Deg$ latitude bins and the $\chi^2$ of data
vs background are calculated. We define\\
\begin{center}
\begin{tabular}{lcl}
	$\chi^2$ &= & $\sum_{i=1,\mu} {(D_i -B_i)^2 / S_i^2}$ \\
        $D_i$ & = & number of events in energy $E$ and latitude bin $i$ \\
        $B_i$ & = & number of events of isotropic background  \\
        $S_i$ & = & standard deviation of $B_i$ \\
	$\mu$ & = & degrees of freedom \\
\end{tabular}
\end{center}
The probability of having a greater or equal $\chi^2$ due to fluctuation in an
isotropic distribution is also computed. A large probability shows that two
distributions are similar to each other and therefore consistent with the
null hypothesis. Conversely, a small probability shows that two
distribution are incompatible. Table 3 lists the $\chi^2$ and
probability $P(>\chi^2,\mu)$.

\placetable{tab3}

	The probability $P(>\chi^2)$ indicates that some energy intervals may 
be inconsistent with the isotropic expectation. To find a functional form of 
this anisotropy, we fit the data to two assumptions. First we look 
for a North-South anisotropy in the latitude $b_i$ (GRAD fit)
\[	R(b_i)=1+f\times b_i \]
Second we look for an excess from the galactic plane using a plane enhancement 
factor (WWFE fit)
\[	R(b_i)=1-f+f\times c\times \Exp{-b_i^2} \]
where $c$ is a normalization constant. Wolfendale and Chi (\cite{WWFE}) claim 
that $c$ is 1.402, however, according to our calculation, $c$ should be 1.437 
(\cite{MAH96}). Here we use $c=1.437$. 
%% 	We also use $c=1.402$ to find the f,
%% 	the result is about 1\% -2\%larger than result use $c=1.437$.

	The fitting is by minimizing  $\chi^2$ defined by
\[	\chi^2=\sum_{i=1,N} {\frac{(D_i - A_i)^2}{E_i^2}}	\]
where
\begin{center}	
  \begin{tabular}{lcl}
	N   &=& number of available bins (bins that have $B_i \not= 0$), \\
	$A_i$ &=& number of expected events in latitude bin $i$,  
		$A_i = B_i \times  R(b_i)$ \\
	$E_i$ &=& error of $A_i$ \\
  \end{tabular}
\end{center}

The fluctuation of the expected number of events, $S_i$, based on the scrambled
 event method, does not follow a Poisson distribution exactly. We find 
that  $S_i$ is approximately 93\% - 95\% of $\sqrt{B_i}$ the expected 
Poisson error. The two can be related by a quadratic form
\[	k \times  S_i^2 + S_i - \sqrt{B_i} =0  	\]
where $k$ depends on whether we use monocular or stereo data and the sky zones. 
When $B_i$ is small, $S_i$ approaches $\sqrt{B_i}$. $E_i$ can be calculated by 
solving
\[	k\times E_i^2 + E_i - \sqrt{A_i} = 0	\]
\[	E_i = \frac{-1 + \sqrt{1+4k\sqrt{A_i}}}{2k}	\]
Figure 2 shows the difference $S_i - \sqrt{B_i}$.
We also show the best fit to the above quadratic form. We use this relation to 
find the error $E_i$ for any given $A_i$.
%	and the difference of the fitted error $E_i - \sqrt{B_i}$.

\placefigure{fig:diff_sd}

	To determine if the data requires a particular functional form (GRAD or
 WWFE), we perform an F-test on the difference of $\chi^2$. 
\[	F = \frac{\chi^2_0 - \chi^2_{min}} {\chi^2_{min}/dof} \]
where 
\begin{center}
  \begin{tabular}{lcl}
	$\chi^2_0$ &=& original $\chi^2$ without functional form ($R(b)=1$) \\
	$\chi^2_{min}$ &=& $\chi^2$ of best fit \\
	dof &=& degrees of freedom \\
  \end{tabular}
\end{center}

The number of degrees of freedom equals N-1 for the full sky zone and the 
supergalactic zone. For the half sky zone, we need to take out one more degree 
of freedom to normalize the total number of events. The probability of the 
F-test $Prob(F;1;dof)$ is then calculated. A small probability suggests the 
functional form is necessary.

	The error of the fit parameter $f$ is calculated by a similar process. 
The upper and lower bounds on $f$ are the values of $f$ at $\chi^2_{min} + 1$. 
The error $df$ is then calculated as one half of the difference of the upper 
and lower bound. 

	The same procedures are applied to the 5,000 simulated data sets. The 
result of theses fits provides a check on the systematic error of the fit 
procedure. The significance of the fit to the data ($f_{data}$) should be 
compared to the mean fitted value ($f_{mean}$) and standard deviation 
($f_{sd}$) of fitted value of all simulated data sets. Then $\sigma$ is 
defined as
\[		\sigma= \frac{f_{data}-f_{mean}} {f_{sd}} \]
A single side Gaussian probability of this $\sigma$ is also calculated 
\[ P_{gauss} =\frac{1}{\sqrt{2\pi}} \int_{\sigma} ^{\infty} 
		\Exp{- \frac{z^2}{2}} dz	\]
	To compare the data with the simulated data sets, a probability of
having a greater fitted value is calculated by
\[	P_{sim}(>f_{data})= \frac{ {\chem{\#\; of\; data\; sets\; that\; 
				generate\; } } f\; >\; f_{data} }
			  {\chem{total\; \#\; of\; data\; sets}}	\]
%%		     # of data sets that generate fitted value > f
%%	P_sim(>f)= --------------------------------------------------
%%			total # of data sets
The total number of simulated data sets is 5,000. Figure 3 shows the histogram
of the WWFE fit parameter $f_E$ for the simulated isotropic data sets. The
distribution of fit parameters can be fitted to a Gaussian form.  This figure
shows the statistical fluctuation of an isotropic background.  The fit
parameter for the real data, $f_{data}$, is shown by the solid arrow.

A small value of probability($P < 0.05$) suggests the fit parameter is 
large compared to the fluctuation of isotropic background. A large 
value ($P > 0.95$) suggests that the fit parameter is small. 

Tables 4 thru 7 list the fit results in galactic 
coordinates. Tables 8 and 9 list 
the fit results in supergalactic coordinates. Figure 4 
thru 9 show the fit results for various conditions.

\placefigure{fig:e2prob}

\notetoeditor{It would be nice if you can place the following pairs of 
table and figure in the same page.}
\placetable{wwfe_all}
\placefigure{fig:wwfe-all}

\placetable{wwfe_half}
\placefigure{fig:wwfe-half}

\placetable{grad_all}
\placefigure{fig:grad-all}

\placetable{grad_half}
\placefigure{fig:grad-half}

\placetable{wwfe_sgb}
\placefigure{fig:wwfe-sgb}

\placetable{grad_sgb}
\placefigure{fig:grad-sgb}

\section{Discussion}
\subsection{Systematic error:}
Due to the limited coverage of galactic latitude, the isotropic background does
 not always produce a null plane enhancement factor or gradient. However, the 
systematic bias is negligible at energies $<3.2EeV$. The systematic bias is 
higher at energy $>3.2EeV$ and the fit parameter may not be the true magnitude 
of the anisotropy. However, the significance and probability are not affected 
by this bias, because the data is compared with simulated data sets. The 
probabilities quoted in the latter part of this article all come from $P_{sim}$.

We also make a study of the systematic bias by reducing the number of events
in each energy bin and redoing the fitting. Figure 10 shows the 
distribution of fit parameters and background as a function of total number of 
events. This study shows that the fit result does not depend strongly of event 
number. However when the number of event falls below 100, the isotropic 
background begins to give $f_E$ significantly different from 0. This shows the 
systematic bias is an effect of low statistics. We suspect the large negative 
$f_E$ reported in some studies (\cite{WWFE}) could also be affected by the 
large bias due to low statistics. 

\placefigure{fig:chk_bkg}

	The other systematic effect is that the error is not Poisson 
distributed. If we use the square root of the expected number $\sqrt{B_i}$ as 
the error, we will over-estimate the error in higher event count regions, 
galactic latitude $-25\Deg$ to $65\Deg$. 
%%	Therefore the fit result will become smaller and the mean fit parameter
%%	will become nonzero. 
By using the fitted error, we improve both the $\chi^2$ value and the accuracy 
of the fit.

\subsection{Anisotropy in galactic coordinates:}
	To search for anisotropy, we first look at the $\chi^2$ between data 
and isotropic background. The $P(>\chi^2,\mu)$ in table 10 shows that 
E2 has $P(>\chi^2,\mu)$ less than 0.20 for all data. This may indicate that the
 data distribution is inconsistent with an isotropic background. On the other 
hand, for E4, all the $P(>\chi^2,\mu)$ are larger than 0.5, This might suggest 
the cosmic ray distribution at E4 is consistent with an isotropic background.

The most significant result is the WWFE fit for E2 ($0.4-1.0EeV$). The 
significance varies from 0.031 to 0.002. Both mono and stereo 
data give the fit plane enhancement factor $f_E$ at about 0.10. The F-test 
probabilities, 0.012 for mono and 0.108 for stereo, also show that such a 
WWFE fit indeed reduces the $\chi^2$. A non-zero $f_E$ is needed to 
fit the data. Finally, the isotropic background has a chance probability 
$P_{sim} = 0.002$ for mono data and 0.016 for stereo data to produced $f_E$ 
larger than the $f_E$ of the data. The gradient fit produces $P_{sim}= 0.731$ to
$0.952$ or $\sim 2\sigma$, which is not sufficiently significant. The data thus
 supports a small anisotropy in E2 ($0.4 -1.0 EeV$). Figure 11 
shows the number of events, event rate, and significance in the E2 bin. 

\placefigure{fig:e2z1nrs}

	Table 10 lists the probability values of all the fits. The 
table shows that most of the fits have similar probability for full sky and 
half sky zones except the WWFE fit to stereo data at E3. Similar result 
can be found for mono and stereo data except for the E1 gradient fit 
and the E4 WWFE fit. 

\placetable{tab9}

	Based on previous FE results on spectrum and composition, we expect
that the anisotropy in the energy range below 3.0 EeV may be different from
that of the higher energy flux. The lower energy data appears to be of a
heavier composition and is likely to be of a galactic origin. We can separate
data into two groups: group 1, E1-E3 Energy $< 3.2EeV$ and group 2, E4-E6
energy $> 3.2EeV$. We use compound probability (\cite{Fisher,Eadie}) to
combine probabilities in table 10.  The results are listed in
table 11.

\begin{equation}
  \begin{array}{lcl}
	p' &=& P(E1)\times P(E2)\times P(E3) \\
	CP &=& p'\times (1.-\log(p')+\frac{1}{2}(\log p')^2)  \\
  \end{array}
\end{equation}

\placetable{tab10}

	Table 11 show that group 1 supports a positive $f_E$ at approximately
$3.2-3.3\sigma$ for monocular data and $1.4-1.9\sigma$ for stereo data.  This
result suggests that cosmic rays with energy$<3.2EeV$ may have a small
anisotropy related to the galactic plane. The stereo data has a less
significant anisotropy because of a very small $f_E$ and large probability at
E1 ($0.2-0.4EeV$). Detailed comparison between mono data and stereo data shows
that there is in fact an excess of events near the galactic center in the mono
data. However, the stereo data acceptance is small in this region due to
smaller zenith angle coverage. For the galactic latitude gradient, the
compound probabilities are in disagreement between monocular data and stereo
data.

	For group 2, both mono and stereo data have no signs of galactic plane
enhancements. The gradient fit shows nothing significant again. Cosmic rays
 with energy$>3.2EeV$ are consistant with an isotropic background. Due to the 
low statistics and large background bias for energy$>10EeV$, we can not prove 
 any anisotropy in this energy region.

\subsection{Anisotropy in supergalactic coordinates:}
	Table 12 lists the probability $P_{sim}$ of mono and stereo 
data in supergalactic coordinates. Although the E2 $f_E$ fit has a 
$P_{sim}=0.9924$, (approximately $-2\sigma$), the event distribution shows that
 the excess of events comes from supergalactic latitude $>+70\Deg$ or 
$<-70\Deg$, i.e. regions where an excess from the galactic plane could be having a strong effect. This supergalactic polar excess is 
not significant either in indiviual bin (E2 mono data $2.38\sigma$) or 
over three energy bins (E1-E3).
Contrary to the Stanev et al. (\cite{Stn95}) result, we do not see evidence of 
anisotropy coming from the supergalactic plane for energies $>10EeV$. However, 
this result is consistent with Kewley et al. (\cite{Kel96}). They did not find 
excess of supergalactic plane in the southern sky either.

\placetable{tab11}

%\subsection{Plane enhancement}
%	The galactic plane enhancement factor had been used to study the 
%anisotropy from galactic plane for almost 14 years. In this study, we see a 
%small anisotropy based on the probability $P_{sim}$. However, the F-test 
%probabilities $P(F)$ are always less significant than $P_{sim}$. This 
%inconsistency may indicate the galactic plane enhancement formula may not be 
%the best fit to the data. In fact, we also see a small galactic latitude 
%gradient at a smaller statistical significance. This also suggest that the 
%galactic plane enhancement formula is not good enough.

\subsection{Comparison with other results}
	Most ground array experiment use harmonic analysis to look for 
anisotropy. However, the Fly's Eye does not have uniform coverage in Right 
Ascension and this  makes harmonic analysis difficult to interpret.

	There had been several reports on the anisotropy toward the galactic 
plane. Gillerman \& Watson (\cite{Gil93}) combined several groups data on 
$f_E$ and calculate the $\chi^2$ for a null fit ($f_E=0$). They report a 
$\chi_{\mu}^2=2.2$ or a probability of 0.9\% that $f_E$ is consistent with zero.
 There may be problems with possible different energy scales between these 
groups of data. In this study, we use the same data set, therefore there is no 
problem with energy cross-calibration. The overall result still supports an 
enhancement from the galactic plane.

Recently, the AGASA group reported a significant first harmonic at energy 
$10^{17.9} - 10^{18.3}eV$, with a chance probability of 0.005\% 
(\cite{Agasa98}). Considering the possible energy scale difference between the 
Fly's Eye and the AGASA, this range may well overlap with our range 
$10^{17.6}-10^{18}eV$. A detailed comparison of the two groups results is 
currently under study.

\section{Summary}
\begin{enumerate}
\item We reexamine both the monocular and stereo Fly's Eye data and look for 
anisotropy related to galactic or supergalactic latitude. Two functional 
forms are studied. The first is the latitude gradient
\[	I(b)=I_0 ( 1+ f \times  b)	\]
the second is plane enhancement factor
\[	I(b)=I_0 ( 1-f +1.437\times f\times \Exp{-b^2}). \]

\item The isotropic background is calculated by scrambling the event arrival 
direction and trigger time. There are 5,000 simulated data sets, each having
the same number of events as the real data. The mean value of these 5,000 sets
are used to represent the distribution expected from isotropy. The standard
deviation of these 5,000 sets is used as the uncertainty in that distribution.

\item We find that the galactic plane enhancement factor $f_E$ is non-zero 
	at the $3.2\sigma$ level for energies $< 3.2EeV$. The chance probability
 	of such an anisotropy existing in these 3 energy bins is less than
 	0.06\%. For energy $> 3.2EeV$, although a negative plane enhancement
 	factor could be possible, the significance is less than $2\sigma$. The
 	galactic latitude gradient is only significant in the energy range $0.4 -1
 	EeV$ where the galactic plane enhancement is strongest too.  Overall,
 	the galactic latitude gradient is not as significant as the galactic
 	plane enhancement factor.  One should bear in mind that the actual
 	form of the anisotropy may be different from either the WWFE or GRAD
 	functions.

\item No significant supergalactic latitude gradient or  supergalactic plane
 	enhancement factor is found. 

\item This analysis supports the the view that the arrival directions of 
	cosmic rays at energies $< 3.2EeV$ are weakly correlated with the
	galactic plane. For energy $>3.2EeV$, no significant anisotropy is
	found.

%\item Implication:
%	Combined this study and composition, spectrum, at energy $<3.2EeV$,
%	Iron dominate the cosmic ray flux and most likely coming from galactic
%	plane. For energy $> 3.2EeV$, proton dominate the flux and the source 
%	probably not in the galactic plane and could be isotropically 
%	distributed.
\end{enumerate}

\acknowledgments
We acknowledge the Department of the Army and the staff of Dugway Proving 
Ground for their cooperation and assistance.
This work has been supported in part by the National Science Foundation (at Utah)  and the U.S. Department of Energy (at Illinois). 

\clearpage

\begin{deluxetable}{ll}
\footnotesize
\tablecaption{Tight cuts used in this study.}\label{tab1}
\tablewidth{0pt}
\tablehead{\colhead{Parameter} & \colhead{Requirement} }
\startdata
        track length            &       $\geq 50\Deg$ \\
        zenith angle            &       $\leq 80\Deg$ \\
        angular uncertainty $\delta$\tablenotemark{a} &  $\leq 15\Deg$ \\
        relative error in energy &      $\leq 3.0$ \\
        relative error in $X_{max}$ &   $\leq 1.0 $ \\
        $X_{max}$ & $300+80\times \log(Energy) \leq X_{max} \leq 
	1100+80\times \log(energy)$ \\
        weather & No frost, moderate or no scattering, $< 1/4$ sky cloudy \\
\enddata
\tablenotetext{a}{angular uncertainty $\delta= \sqrt {d\theta^2 + 
	(\sin\theta\times d\phi)^2 }$}
\end{deluxetable}

\clearpage

\begin{deluxetable}{lcrrrr}
\footnotesize
\tablecaption{Number of events and mean energy of events that pass the tight cuts}
        \label{tab2}
\tablewidth{0pt}
\tablehead{\multicolumn{2}{c}{Energy} & \multicolumn{2}{c}{Stereo} & 
\multicolumn{2}{c}{Mono} \\ \multicolumn{2}{c}{$(EeV)$} &
\colhead{Num} & \colhead{$<E>$} & \colhead{Num} & \colhead{$<E>$} }
\startdata
E1 & $0.2 - 0.4$ &     2709 &    0.289 &         5183 &    0.287 \\
E2 & $0.4 - 1.0$ &     2402 &    0.619 &         4815 &    0.621 \\
E3 & $1.0 - 3.2$ &     1141 &    1.613 &         2465 &    1.658 \\
E4 & $3.2 -10.0$ &      191 &    5.912 &          597 &    5.357 \\
E5 & $> 10.0$    &       33 &   16.366 &          169 &   22.806 \\
E6 & $> 32.0$    &        2 &   41.786 &           17 &   78.282 \\
\tableline
\multicolumn{2}{c}{Total} & 6476 & & 13229& \\
\enddata
\end{deluxetable}

\clearpage

\begin{deluxetable}{ccrrrlrrrl}
\footnotesize
\tablecaption{The $\chi^2$ for data $D_i$ and isotropic background $B_i$.}
        \label{tab3}
\tablewidth{0pt}
\tablehead{\colhead{ } & \colhead{ } & 
	\colhead{Num. of} & \multicolumn{3}{c}{Mono} & 
	\colhead{Num. of} & \multicolumn{3}{c}{Mono} \\
	\colhead{Zone} & \colhead{Energy} 
	& \colhead{events} & \colhead{$\mu$} & \colhead{$\chi_{\mu}^2$} &
	\colhead{$P(>\chi^2,\mu)$}
	& \colhead{events} & \colhead{$\mu$} & \colhead{$\chi_{\mu}^2$} &
	\colhead{$P(>\chi^2,\mu)$} }
\startdata
\multicolumn{10}{l}{Galactic latitude:} \\
Full sky & E1 &   5183 &  17 & 1.105 & 0.341 &  2709 &  17 & 0.715 & 0.790 \\
        & E2 &   4815 &  17 & 1.355 & 0.148 &   2402 &  17 & 1.673 & 0.040 \\
        & E3 &   2465 &  17 & 1.290 & 0.187 &   1141 &  17 & 0.830 & 0.659 \\
        & E4 &    597 &  17 & 0.631 & 0.870 &    191 &  17 & 0.910 & 0.561 \\
        & E5 &    169 &  17 & 0.988 & 0.468 &     33 &  16 & 0.714 & 0.791 \\
        & E6 &     17 &  16 & 0.923 & 0.545 \\
\tableline
Half sky & E1 &  4553 &  16 & 1.312 & 0.179 &   2289 &  16 & 0.738 & 0.757 \\
        & E2  &  4092 &  16 & 1.430 & 0.117 &   2031 &  16 & 1.279 & 0.200 \\
        & E3  &  2040 &  16 & 1.071 & 0.377 &    972 &  16 & 0.655 & 0.840 \\
        & E4  &   470 &  16 & 0.761 & 0.731 &    152 &  16 & 0.820 & 0.664 \\
        & E5  &   130 &  16 & 0.928 & 0.535 &     24 &  15 & 0.631 & 0.862 \\
        & E6  &    14 &  15 & 1.000 & 0.453 \\
\tableline
\multicolumn{10}{l}{Supergalactic latitude:}\\
        & E1  &  5183 &  17 & 0.919 & 0.551 &   2709 &  17 & 1.780 & 0.025 \\
        & E2  &  4815 &  17 & 1.567 & 0.064 &   2402 &  17 & 0.601 & 0.894 \\
        & E3  &  2465 &  17 & 1.313 & 0.173 &   1141 &  17 & 0.620 & 0.879 \\
        & E4  &   597 &  17 & 1.090 & 0.356 &    191 &  17 & 0.757 & 0.745 \\
        & E5  &   169 &  17 & 1.278 & 0.195 &     33 &  17 & 0.915 & 0.556 \\
        & E6  &    17 &  17 & 0.727 & 0.778 \\
\enddata
\end{deluxetable}

\clearpage

\begin{deluxetable}{lrrrrrrrrrr}
\footnotesize
\tablecaption{Results for WWFE fitting using the full sky data. P(F) is the
        F-test probability $Prob(F;1;dof)$.}
        \label{wwfe_all}
\tablewidth{0pt}
\tablehead{\colhead{E} & \colhead{Data} & \colhead{$f$} & \colhead{$df$} & 
\colhead{$\chi^2_{min}$} & \colhead{F} & \colhead{P(F)} & \colhead{$f_{mean}$} 
& \colhead{$f_{SD}$} & \colhead{$P_{gauss}$} & \colhead{$P_{sim}$} }
\startdata
E1 &mono& 0.067 & 0.035 & 0.943 & 3.922 & 0.065 &-0.003 & 0.037 & 0.030 & 0.029 \\
 &stereo& 0.016 & 0.047 & 0.753 & 0.151 & 0.702 &-0.003 & 0.050 & 0.352 & 0.354 \\
E2 &mono& 0.104 & 0.036 & 0.955 & 8.120 & 0.012 &-0.002 & 0.038 & 0.003 & 0.002 \\
 &stereo& 0.109 & 0.052 & 1.505 & 2.896 & 0.108 &-0.004 & 0.054 & 0.018 & 0.016 \\
E3 &mono& 0.063 & 0.052 & 1.279 & 1.145 & 0.300 &-0.004 & 0.056 & 0.116 & 0.116 \\
 &stereo& 0.079 & 0.080 & 0.823 & 1.152 & 0.299 &-0.008 & 0.077 & 0.128 & 0.132 \\
E4 &mono&-0.072 & 0.113 & 0.645 & 0.635 & 0.437 &-0.016 & 0.111 & 0.694 & 0.705 \\
 &stereo&-0.049 & 0.183 & 0.963 & 0.075 & 0.787 &-0.044 & 0.195 & 0.511 & 0.514 \\
E5 &mono&-0.264 & 0.218 & 0.954 & 1.616 & 0.222 &-0.061 & 0.225 & 0.817 & 0.818 \\
E5 &ste &-0.576 & 0.402 & 0.631 & 3.235 & 0.091 &-0.232 & 0.454 & 0.776 & 0.774 \\
E6 &mono&-1.263 & 0.447 & 0.618 & 9.379 & 0.007 &-0.243 & 0.580 & 0.961 & 0.959 \\
\enddata
\end{deluxetable}

\clearpage

\begin{deluxetable}{lrrrrrrrrrr}
\footnotesize
\tablecaption{Results for WWFE fitting using the half sky data.}
        \label{wwfe_half}
\tablewidth{0pt}
\tablehead{\colhead{E} & \colhead{Data} & \colhead{$f$} & \colhead{$df$} & 
\colhead{$\chi^2_{min}$} & \colhead{F} & \colhead{P(F)} & \colhead{$f_{mean}$} 
& \colhead{$f_{SD}$} & \colhead{$P_{gauss}$} & \colhead{$P_{sim}$} }
\startdata
E1 &mono& 0.096 & 0.040 & 1.016 & 5.659 & 0.031 &-0.003 & 0.042 & 0.009 & 0.008 \\
 &stereo& 0.045 & 0.056 & 0.745 & 0.864 & 0.367 &-0.006 & 0.058 & 0.194 & 0.196 \\
E2 &mono& 0.126 & 0.043 & 0.972 & 8.539 & 0.011 &-0.003 & 0.045 & 0.002 & 0.002 \\
 &stereo& 0.113 & 0.061 & 1.144 & 2.879 & 0.110 &-0.006 & 0.063 & 0.029 & 0.031 \\
E3 &mono& 0.024 & 0.060 & 1.132 & 0.137 & 0.716 &-0.007 & 0.064 & 0.318 & 0.317 \\
 &stereo&-0.039 & 0.093 & 0.687 & 0.252 & 0.623 &-0.014 & 0.091 & 0.608 & 0.615 \\
E4 &mono&-0.156 & 0.130 & 0.715 & 2.042 & 0.173 &-0.026 & 0.131 & 0.839 & 0.844 \\
 &stereo& 0.098 & 0.215 & 0.862 & 0.233 & 0.636 &-0.073 & 0.228 & 0.227 & 0.226 \\
E5 &mono&-0.261 & 0.252 & 0.915 & 1.230 & 0.285 &-0.094 & 0.261 & 0.739 & 0.737 \\
 &stereo&-0.486 & 0.454 & 0.596 & 1.941 & 0.184 &-0.319 & 0.509 & 0.629 & 0.627 \\
E6 &mono&-1.448 & 0.428 & 0.608 &11.313 & 0.004 &-0.323 & 0.668 & 0.954 & 0.959 \\
\enddata
\end{deluxetable}

\clearpage

\begin{deluxetable}{lrrrrrrrrrr}
\footnotesize
\tablecaption{Results for gradient fitting using the full sky data.}
        \label{grad_all}
\tablewidth{0pt}
\tablehead{\colhead{E} & \colhead{Data} & \colhead{$f$} & \colhead{$df$} & 
\colhead{$\chi^2_{min}$} & \colhead{F} & \colhead{P(F)} & \colhead{$f_{mean}$} 
& \colhead{$f_{SD}$} & \colhead{$P_{gauss}$} & \colhead{$P_{sim}$} }
\startdata
E1 &Mono&-0.050 & 0.021 & 0.832 & 6.569 & 0.021 &-0.001 & 0.022 & 0.987 & 0.985 \\
 &stereo& 0.040 & 0.029 & 0.638 & 3.077 & 0.099 &-0.001 & 0.028 & 0.074 & 0.072 \\
E2 &Mono&-0.021 & 0.023 & 1.385 & 0.628 & 0.440 &-0.001 & 0.023 & 0.811 & 0.803 \\
 &stereo&-0.053 & 0.030 & 1.582 & 1.975 & 0.179 &-0.001 & 0.031 & 0.955 & 0.952 \\
E3 &Mono&-0.052 & 0.030 & 1.188 & 2.458 & 0.136 &-0.001 & 0.031 & 0.947 & 0.947 \\
 &stereo&-0.098 & 0.043 & 0.576 & 8.507 & 0.010 &-0.002 & 0.045 & 0.984 & 0.986 \\
E4 &Mono& 0.072 & 0.064 & 0.589 & 2.229 & 0.155 &-0.005 & 0.062 & 0.107 & 0.105 \\
 &stereo& 0.066 & 0.105 & 0.943 & 0.417 & 0.528 &-0.012 & 0.104 & 0.228 & 0.225 \\
E5 &Mono&-0.033 & 0.113 & 1.045 & 0.082 & 0.779 &-0.016 & 0.125 & 0.555 & 0.547 \\
 &stereo& 0.134 & 0.204 & 0.733 & 0.565 & 0.463 &-0.021 & 0.245 & 0.263 & 0.257 \\
E6 &Mono&-0.114 & 0.191 & 0.961 & 0.341 & 0.567 &-0.051 & 0.325 & 0.576 & 0.544 \\
\enddata
\end{deluxetable}

\clearpage

\begin{deluxetable}{lrrrrrrrrrr}
\footnotesize
\tablecaption{Results for gradient fit using the half sky data.}
        \label{grad_half}
\tablewidth{0pt}
\tablehead{\colhead{E} & \colhead{Data} & \colhead{$f$} & \colhead{$df$} & 
\colhead{$\chi^2_{min}$} & \colhead{F} & \colhead{P(F)} & \colhead{$f_{mean}$} 
& \colhead{$f_{SD}$} & \colhead{$P_{gauss}$} & \colhead{$P_{sim}$} }
\startdata
E1 &Mono&-0.066 & 0.024 & 0.897 & 8.411 & 0.011 &-0.001 & 0.024 & 0.997 & 0.997 \\
 &stereo& 0.028 & 0.034 & 0.742 & 0.928 & 0.351 &-0.001 & 0.033 & 0.193 & 0.191 \\
E2 &Mono&-0.017 & 0.026 & 1.497 & 0.286 & 0.600 &-0.001 & 0.026 & 0.738 & 0.731 \\
 &stereo&-0.050 & 0.035 & 1.228 & 1.662 & 0.217 &-0.001 & 0.035 & 0.919 & 0.920 \\
E3 &Mono&-0.026 & 0.034 & 1.106 & 0.491 & 0.494 &-0.001 & 0.036 & 0.758 & 0.755 \\
 &stereo&-0.053 & 0.049 & 0.622 & 1.836 & 0.196 &-0.001 & 0.052 & 0.841 & 0.843 \\
E4 &Mono& 0.058 & 0.069 & 0.766 & 0.898 & 0.358 &-0.003 & 0.072 & 0.198 & 0.197 \\
 &stereo& 0.027 & 0.127 & 0.872 & 0.054 & 0.819 &-0.002 & 0.120 & 0.403 & 0.398 \\
E5 &Mono& 0.048 & 0.144 & 0.983 & 0.112 & 0.743 &-0.011 & 0.143 & 0.339 & 0.339 \\
 &stereo&-0.058 & 0.267 & 0.670 & 0.071 & 0.793 & 0.017 & 0.281 & 0.606 & 0.597 \\
E6 &Mono&-0.133 & 0.215 & 1.042 & 0.352 & 0.562 &-0.028 & 0.373 & 0.611 & 0.594 \\
\enddata
\end{deluxetable}

\clearpage

\begin{deluxetable}{lrrrrrrrrrr}
\footnotesize
\tablecaption{Results for WWFE fitting using the supergalactic latitude.}
        \label{wwfe_sgb}
\tablewidth{0pt}
\tablehead{\colhead{E} & \colhead{Data} & \colhead{$f$} & \colhead{$df$} & 
\colhead{$\chi^2_{min}$} & \colhead{F} & \colhead{P(F)} & \colhead{$f_{mean}$} 
& \colhead{$f_{SD}$} & \colhead{$P_{gauss}$} & \colhead{$P_{sim}$} }
\startdata
E1 & mono&-0.004 & 0.038 & 0.975 & 0.013 & 0.910 &-0.002 & 0.039 & 0.518 & 0.525 \\
  &stereo &-0.052 & 0.050 & 1.824 & 0.592 & 0.453 &-0.004 & 0.052 & 0.824 & 0.827 \\
E2 & mono&-0.095 & 0.038 & 1.247 & 5.368 & 0.034 &-0.002 & 0.038 & 0.992 & 0.991 \\
  &stereo &-0.051 & 0.052 & 0.580 & 1.634 & 0.219 &-0.004 & 0.054 & 0.807 & 0.808 \\
E3 & mono&-0.019 & 0.052 & 1.387 & 0.094 & 0.763 &-0.004 & 0.054 & 0.611 & 0.623 \\
  &stereo &-0.066 & 0.078 & 0.613 & 1.204 & 0.289 &-0.008 & 0.077 & 0.775 & 0.770 \\
E4 & mono&-0.012 & 0.108 & 1.157 & 0.011 & 0.916 &-0.015 & 0.108 & 0.488 & 0.487 \\
  &stereo & 0.113 & 0.178 & 0.780 & 0.495 & 0.492 &-0.039 & 0.185 & 0.206 & 0.204 \\
E5 & mono& 0.075 & 0.203 & 1.350 & 0.101 & 0.755 &-0.059 & 0.219 & 0.271 & 0.281 \\
  &stereo &-0.124 & 0.322 & 0.963 & 0.157 & 0.697 &-0.206 & 0.430 & 0.425 & 0.439 \\
E6 & mono&-0.208 & 0.456 & 0.759 & 0.281 & 0.603 &-0.239 & 0.569 & 0.478 & 0.475 \\
\enddata
\end{deluxetable}

\clearpage

\begin{deluxetable}{lrrrrrrrrrr}
\footnotesize
\tablecaption{Results for gradient fit using the supergalactic latitude.}
        \label{grad_sgb}
\tablewidth{0pt}
\tablehead{\colhead{E} & \colhead{Data} & \colhead{$f$} & \colhead{$df$} & 
\colhead{$\chi^2_{min}$} & \colhead{F} & \colhead{P(F)} & \colhead{$f_{mean}$} 
& \colhead{$f_{SD}$} & \colhead{$P_{gauss}$} & \colhead{$P_{sim}$} }
\startdata
E1 &mono& 0.001 & 0.021 & 0.976 & 0.003 & 0.954 &-0.001 & 0.023 & 0.471 & 0.460 \\
 &stereo &-0.001 & 0.028 & 1.891 & 0.001 & 0.974 &-0.001 & 0.030 & 0.500 & 0.489 \\
E2 &mono& 0.005 & 0.021 & 1.661 & 0.033 & 0.858 &-0.001 & 0.022 & 0.398 & 0.394 \\
 &stereo & 0.017 & 0.030 & 0.617 & 0.557 & 0.466 &-0.001 & 0.031 & 0.280 & 0.276 \\
E3 &mono&-0.047 & 0.029 & 1.235 & 2.071 & 0.169 &-0.001 & 0.031 & 0.928 & 0.924 \\
 &stereo & 0.000 & 0.043 & 0.659 & 0.000 & 1.000 &-0.002 & 0.044 & 0.481 & 0.472 \\
E4 &mono& 0.010 & 0.061 & 1.156 & 0.024 & 0.880 &-0.004 & 0.061 & 0.408 & 0.400 \\
 &stereo &-0.065 & 0.103 & 0.780 & 0.493 & 0.493 &-0.106 & 0.102 & 0.704 & 0.700 \\
E5 &mono&-0.273 & 0.119 & 1.048 & 4.736 & 0.045 &-0.014 & 0.126 & 0.981 & 0.981 \\
 &stereo & 0.168 & 0.265 & 0.945 & 0.454 & 0.510 &-0.039 & 0.237 & 0.191 & 0.185 \\
E6 &mono&-0.178 & 0.250 & 0.742 & 0.662 & 0.428 &-0.045 & 0.320 & 0.661 & 0.638 \\
\enddata
\end{deluxetable}

\clearpage

\begin{deluxetable}{crrrrrrrr}
\footnotesize
\tablecaption{The probability $P_{sim}(>f_{data})$ for all fits.}
        \label{tab9}
\tablewidth{0pt}
\tablehead{\colhead{ } & \multicolumn{4}{c}{WWFE} &\multicolumn{4}{c}{GRAD} \\
\colhead{ } & \multicolumn{2}{c}{Mono} &\multicolumn{2}{c}{Stereo} \\
\colhead{Energy} & \colhead{Full} & \colhead{Half} & \colhead{Full} & 
\colhead{Half} & \colhead{Full} & \colhead{Half} & \colhead{Full} & 
\colhead{Half} }
\startdata
E1 & 0.0292 & 0.0080 & 0.3544 & 0.1962 & 0.9852 & 0.9966 & 0.0722 & 0.1910\\
E2 & 0.0022 & 0.0022 & 0.0162 & 0.0312 & 0.8028 & 0.7308 & 0.9524 & 0.9196\\
E3 & 0.1162 & 0.3166 & 0.1316 & 0.6148 & 0.9472 & 0.7550 & 0.9860 & 0.8432\\
E4 & 0.7048 & 0.8438 & 0.5142 & 0.2258 & 0.1046 & 0.1968 & 0.2254 & 0.3976\\
E5 & 0.8178 & 0.7372 & 0.7744 & 0.6266 & 0.5474 & 0.3386 & 0.2574 & 0.5972\\
E6 & 0.9590 & 0.9590 &        &        & 0.5444 & 0.5942 \\
\enddata
\end{deluxetable}

\clearpage

\begin{deluxetable}{rrrrrrrrr}
\footnotesize
\tablecaption{The compound probability and sigma of all the fits.}
        \label{tab10}
\tablewidth{0pt}
\tablehead{\colhead{} & \multicolumn{4}{c}{Mono} &\multicolumn{4}{c}{Stereo} \\
\colhead{} &\multicolumn{2}{c}{Full} & \multicolumn{2}{c}{Half}  & 
\multicolumn{2}{c}{Full} & \multicolumn{2}{c}{Half} \\
\colhead{} & \colhead{CP} & \colhead{sig} & \colhead{CP} & \colhead{sig} & 
\colhead{CP} & \colhead{sig} & \colhead{CP} & \colhead{sig}}
\startdata
& \multicolumn{4}{l}{E1-E3} & \multicolumn{4}{l}{E1-E3} \\
WWFE CP & 0.0006 & 3.2315 & 0.0005 & 3.3016 & 0.0257 & 1.9481 & 0.0834 & 1.3825 \\
GRAD CP & 0.9968 &-2.7227 & 0.9771 &-1.9968 & 0.4958 & 0.0106 & 0.7011 &-0.5275 \\
\tableline
 & \multicolumn{4}{l}{E4-E6} & \multicolumn{4}{l}{E4-E5} \\
WWFE CP & 0.9776 &-2.0063 & 0.9843 &-2.1525 & 0.7649 &-0.7220 & 0.4182 & 0.2066 \\
GRAD CP & 0.3268 & 0.4489 & 0.3739 & 0.3216 & 0.2232 & 0.7615 & 0.5789 &-0.1989 \\
\enddata
\end{deluxetable}

\clearpage

\begin{deluxetable}{crrrr}
\footnotesize
\tablecaption{The probability $P_{sim}(>f_{data})$ for all the fits in 
	supergalactic latitude.}
        \label{tab11}
\tablewidth{0pt}
\tablehead{\colhead{Fit} &\multicolumn{2}{c}{WWFE} &\multicolumn{2}{c}{GRAD} \\
\colhead{Energy} & \colhead{Mono} & \colhead{Stereo} & \colhead{Mono} & 
\colhead{Stereo} }
\startdata
E1 &0.5246 &0.8274 &0.4602 &0.4890 \\
E2 &0.9914 &0.8076 &0.3936 &0.2762 \\
E3 &0.6234 &0.7696 &0.9240 &0.4720 \\
E4 &0.4866 &0.2040 &0.4002 &0.7000 \\
E5 &0.2806 &0.4386 &0.9810 &0.1846 \\
E6 &0.4748 &       &0.6382 &       \\
\tableline
E1-E3 CP&0.8951 &0.9700 &0.7340 &0.4808 \\
Significance & $1.25\sigma$ & $1.88\sigma$ & $0.63\sigma$ & $0.05\sigma$ \\
\enddata
\end{deluxetable}
% This is the last table for this paper
\clearpage

\clearpage

\figcaption[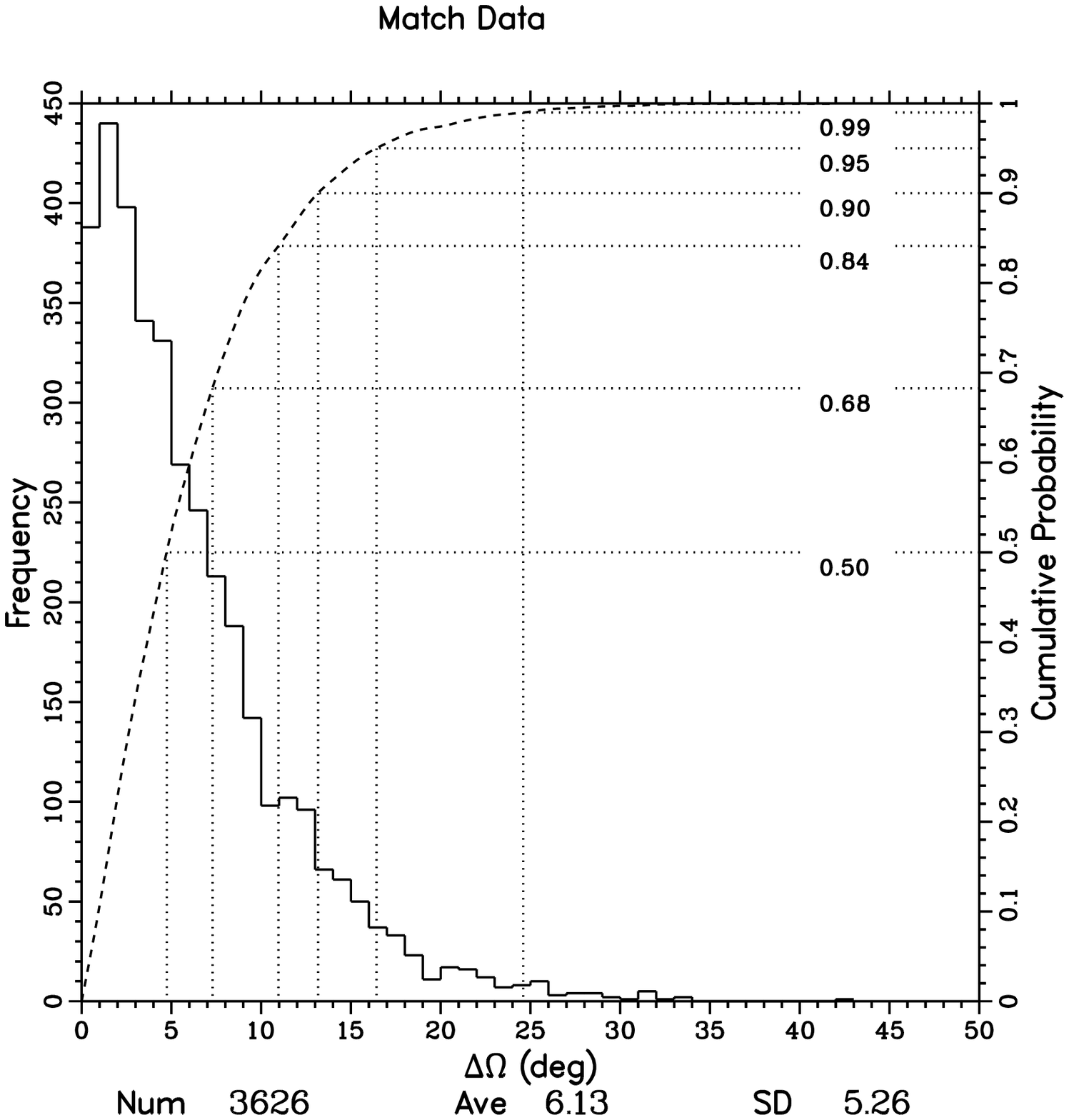]{Angular resolution of common events in both mono 
	and stereo data. The $\Delta\Omega$ is the space angle between arrival 
	directions reconstructed by monocular and stereo methods.}
  \label{fig:ang_res_match}

\figcaption[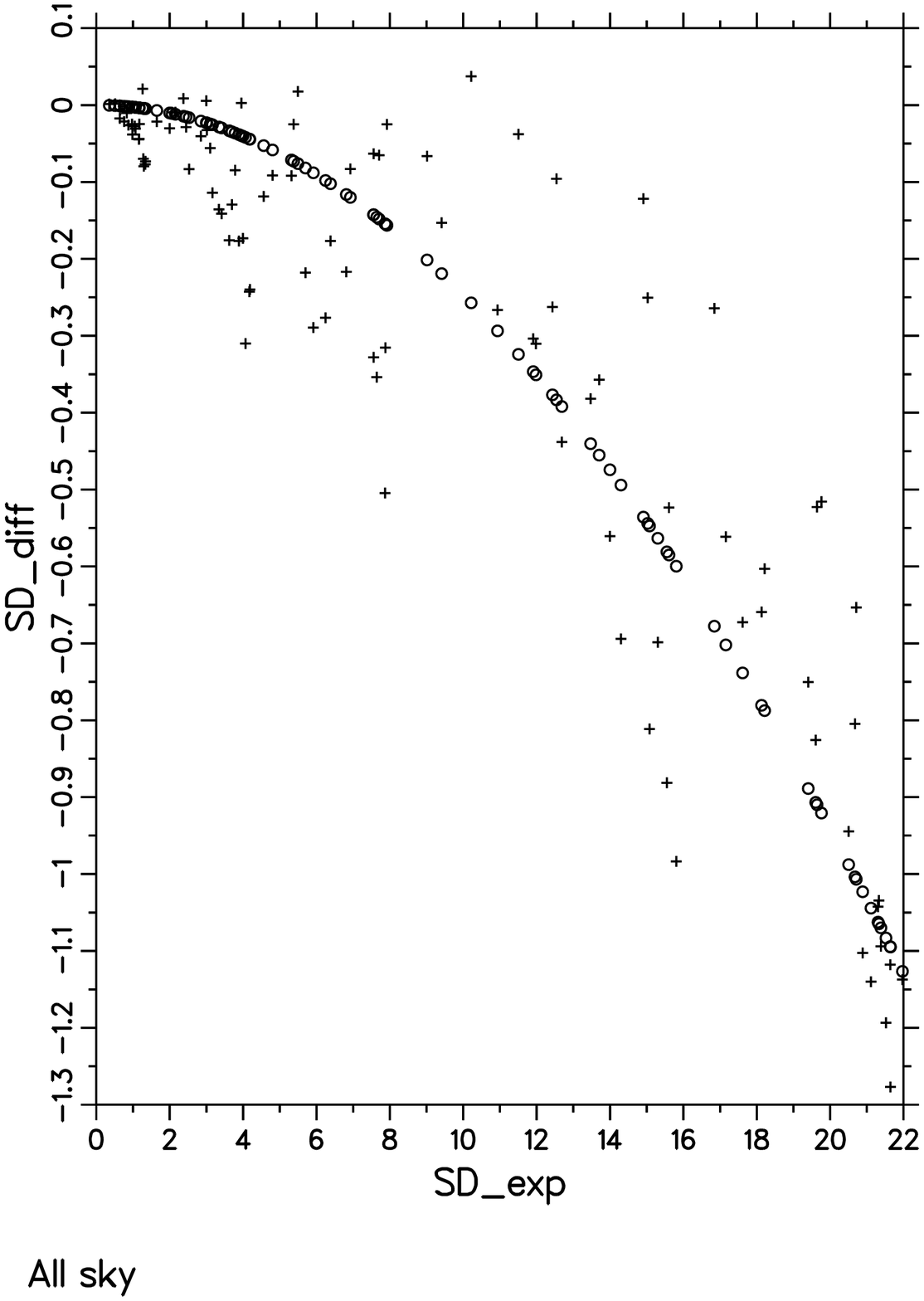]{The X axis is $\sqrt{B_{i}}$ the expected error
         from Poisson distribution. The Y axis is the difference between
        the observed scrambled event method standard deviation and the expected Poisson error $S_i - \sqrt{B_{i}}$, shown as crosses, and the  fitted
        error difference $E_{i} - \sqrt{B_{i}}$, shown as circles.}
	\label{fig:diff_sd}

\figcaption[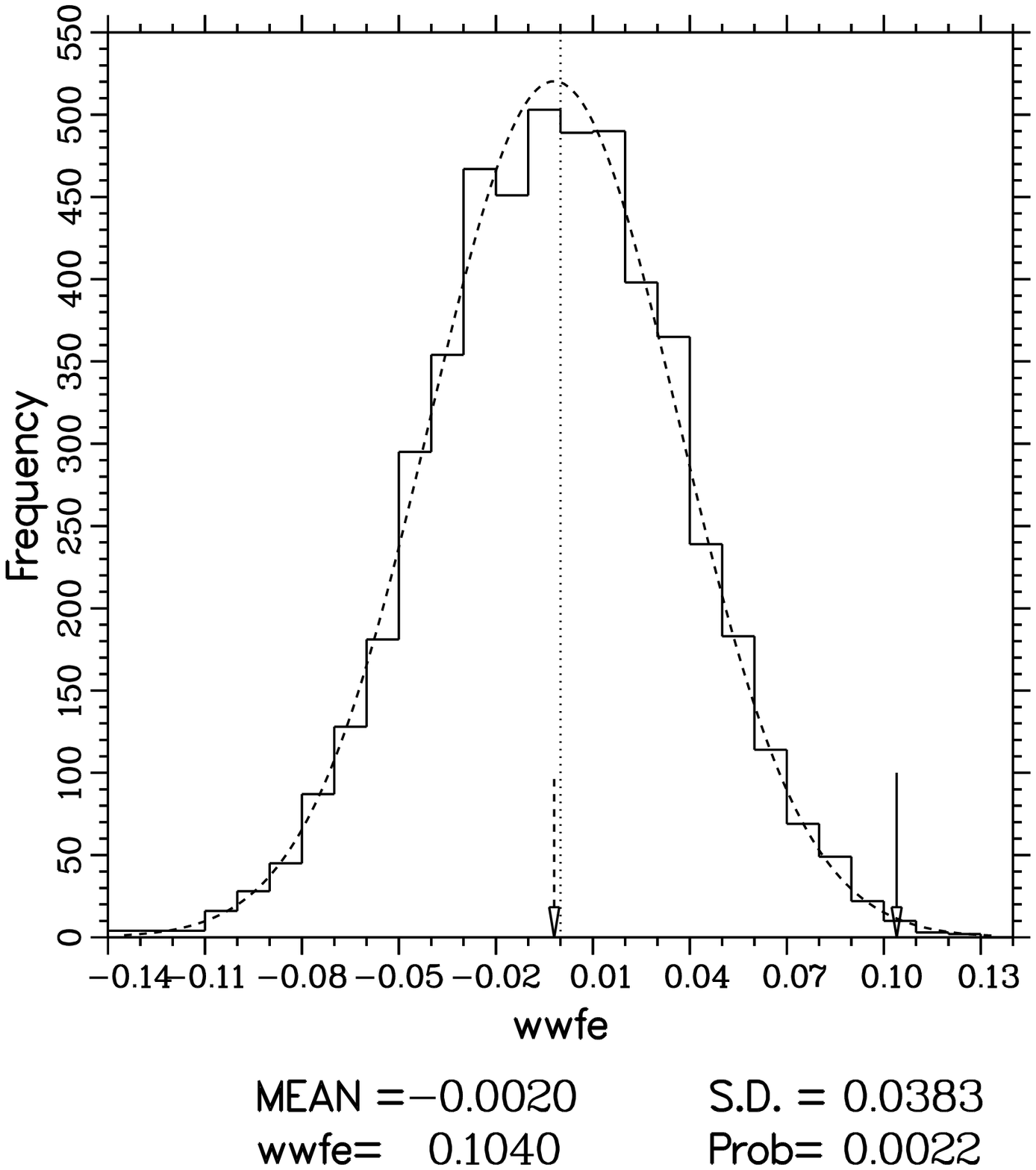]{Histogram of the WWFE fit parameter for monocular data 
	using the full sky at energy of 0.4-1.0EeV. The probability $P_{sim}$ 
	is the histogram area from the fit parameter for the data ($f_{data}$ 
	solid arrow) to the right end of axis. Similarly, the $P_{gauss}$ is 
	the area of the fitted Gaussian form from the fit to the right end of 
	the axis.}
  \label{fig:e2prob}

\figcaption[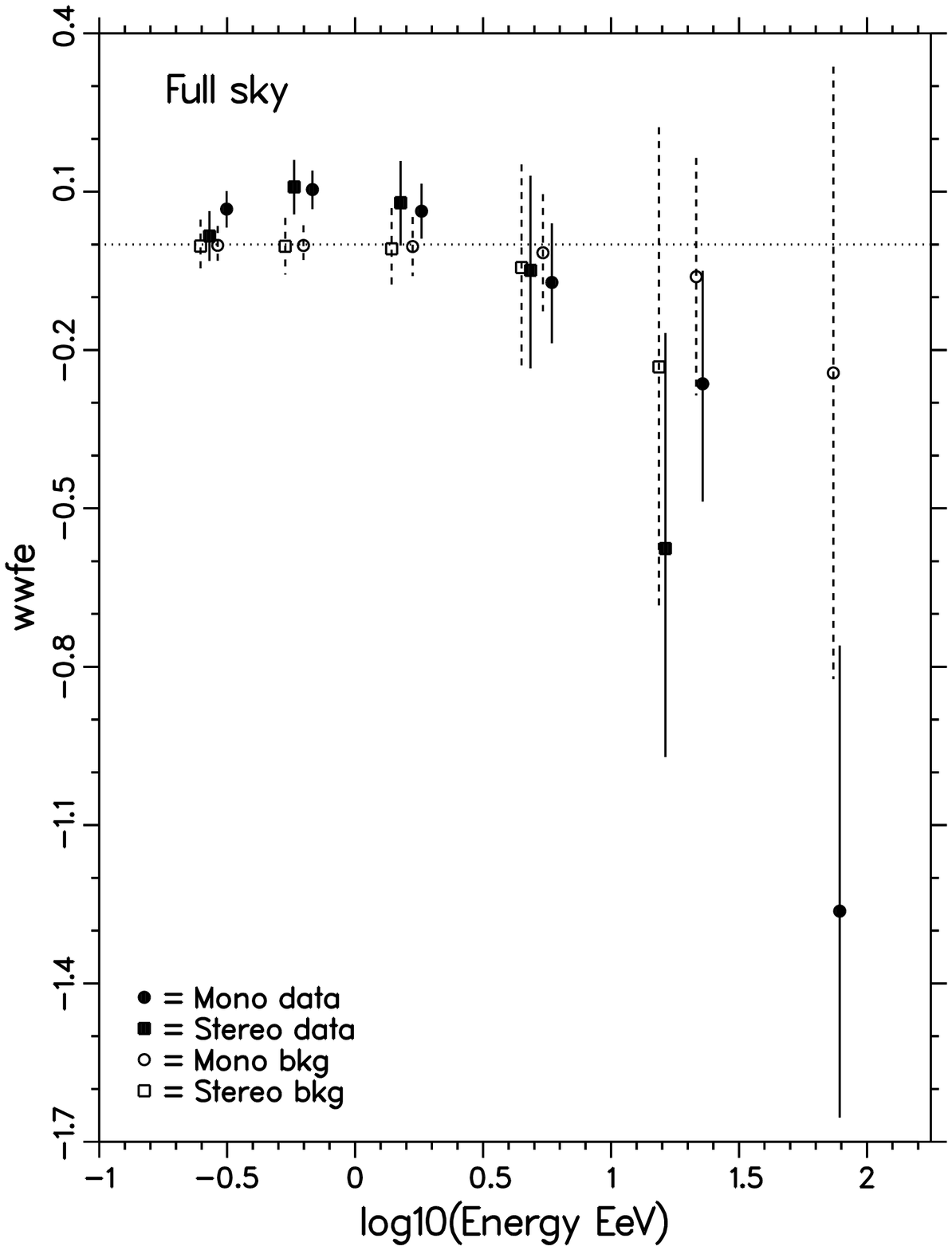]{The WWFE fit using the full sky data.}
  \label{fig:wwfe-all}

\figcaption[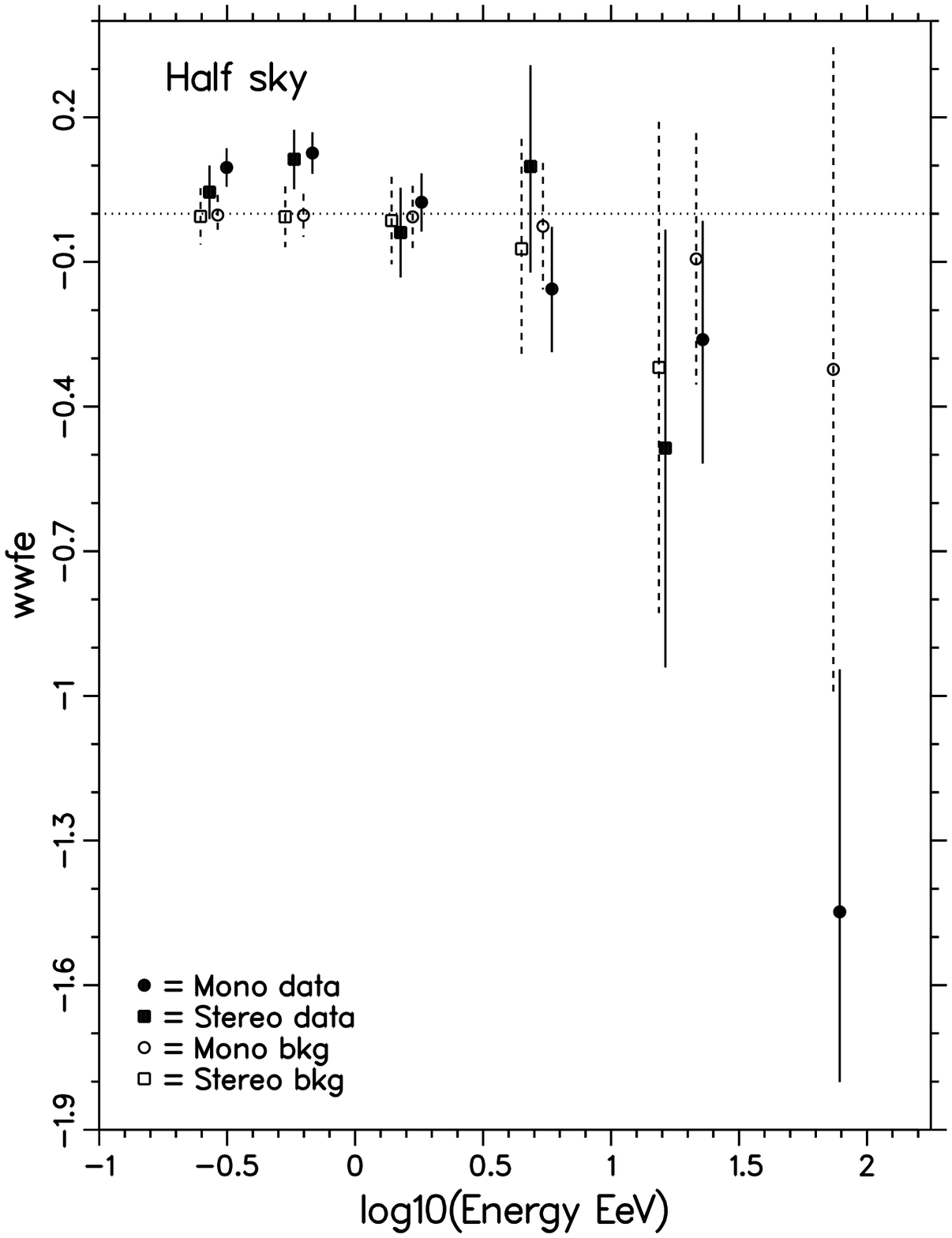]{The WWFE fit using the half sky data.}
  \label{fig:wwfe-half}

\figcaption[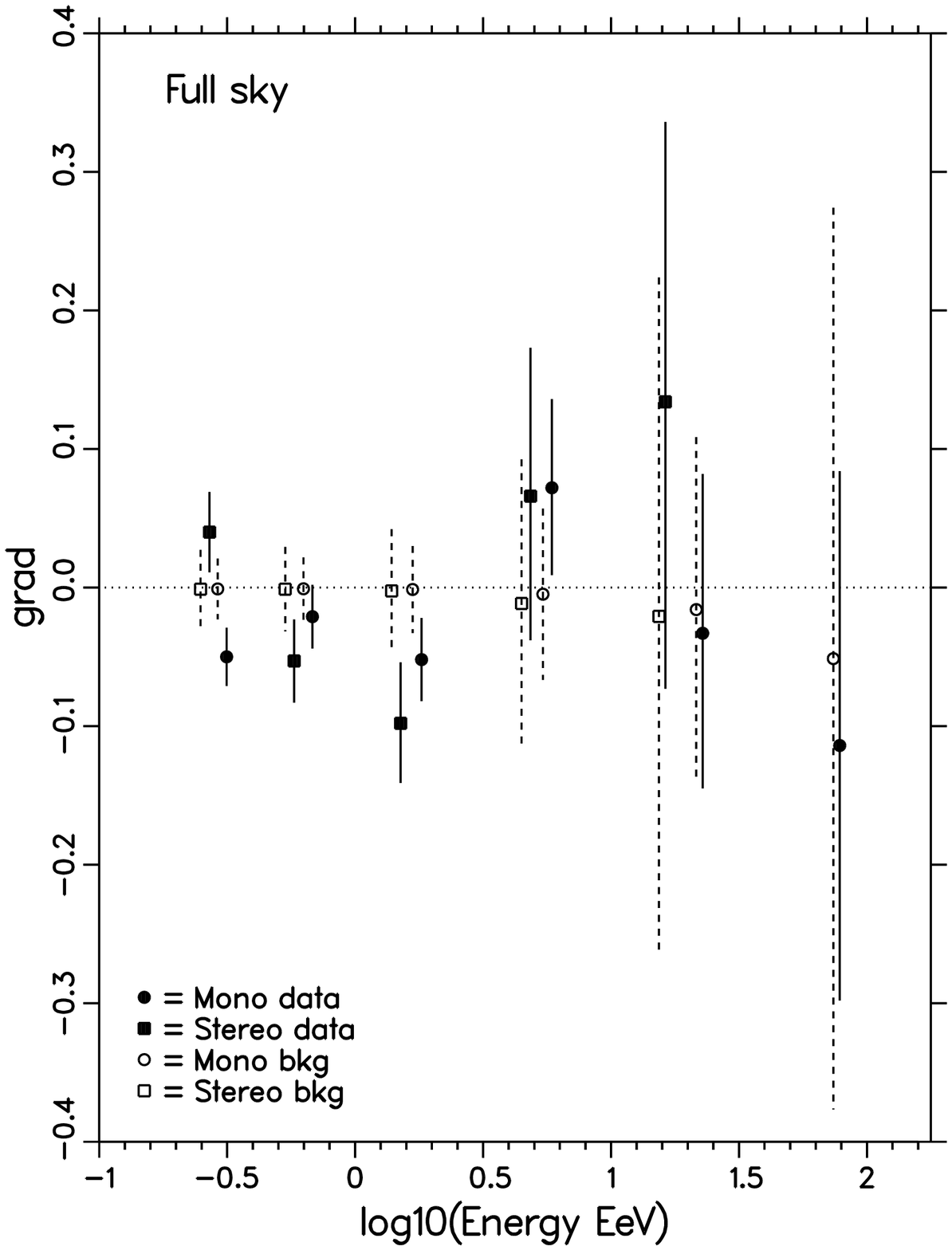]{The gradient fit using the full sky data.}
  \label{fig:grad-all}

\figcaption[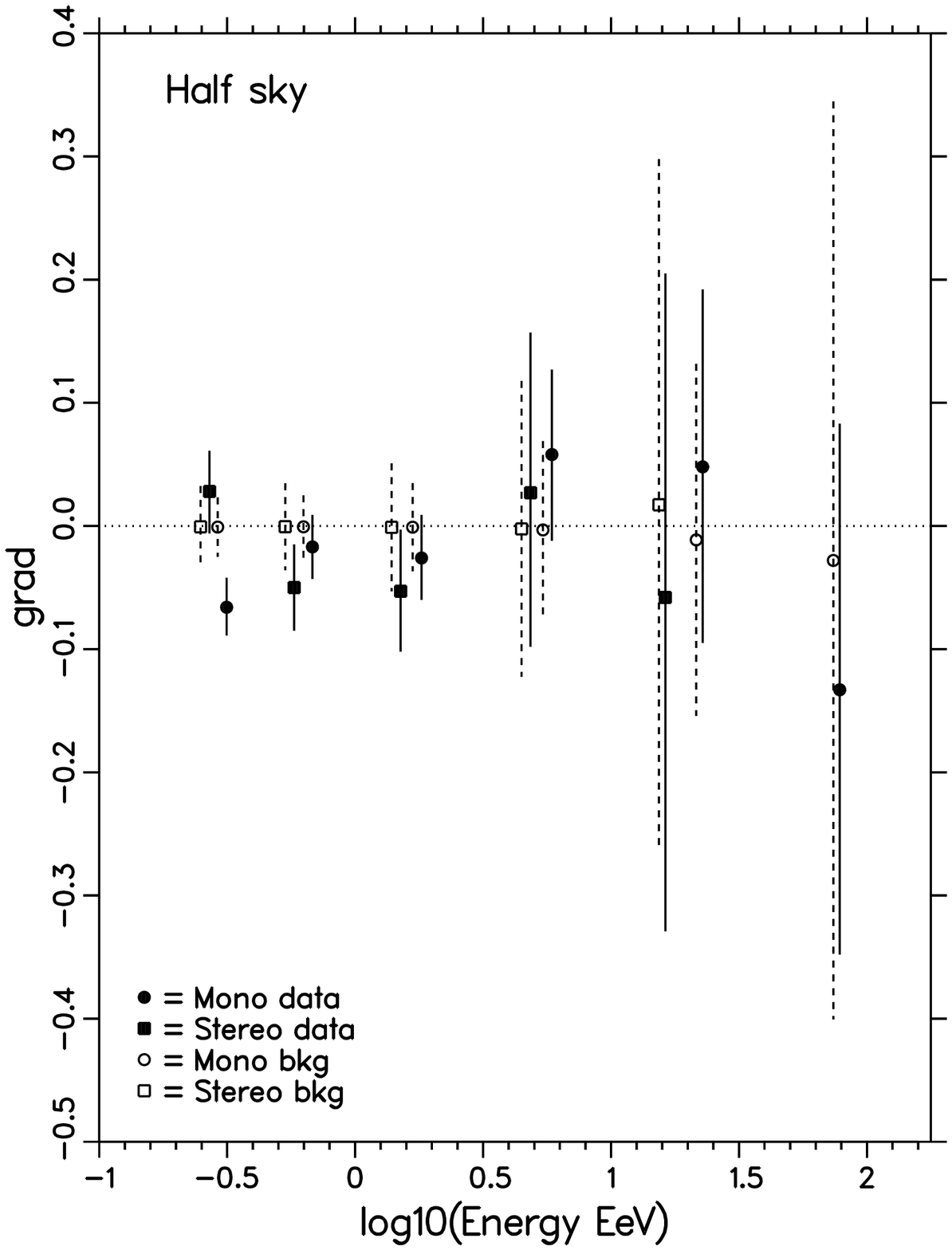]{The gradient fit using the half sky data.}
  \label{fig:grad-half}

\figcaption[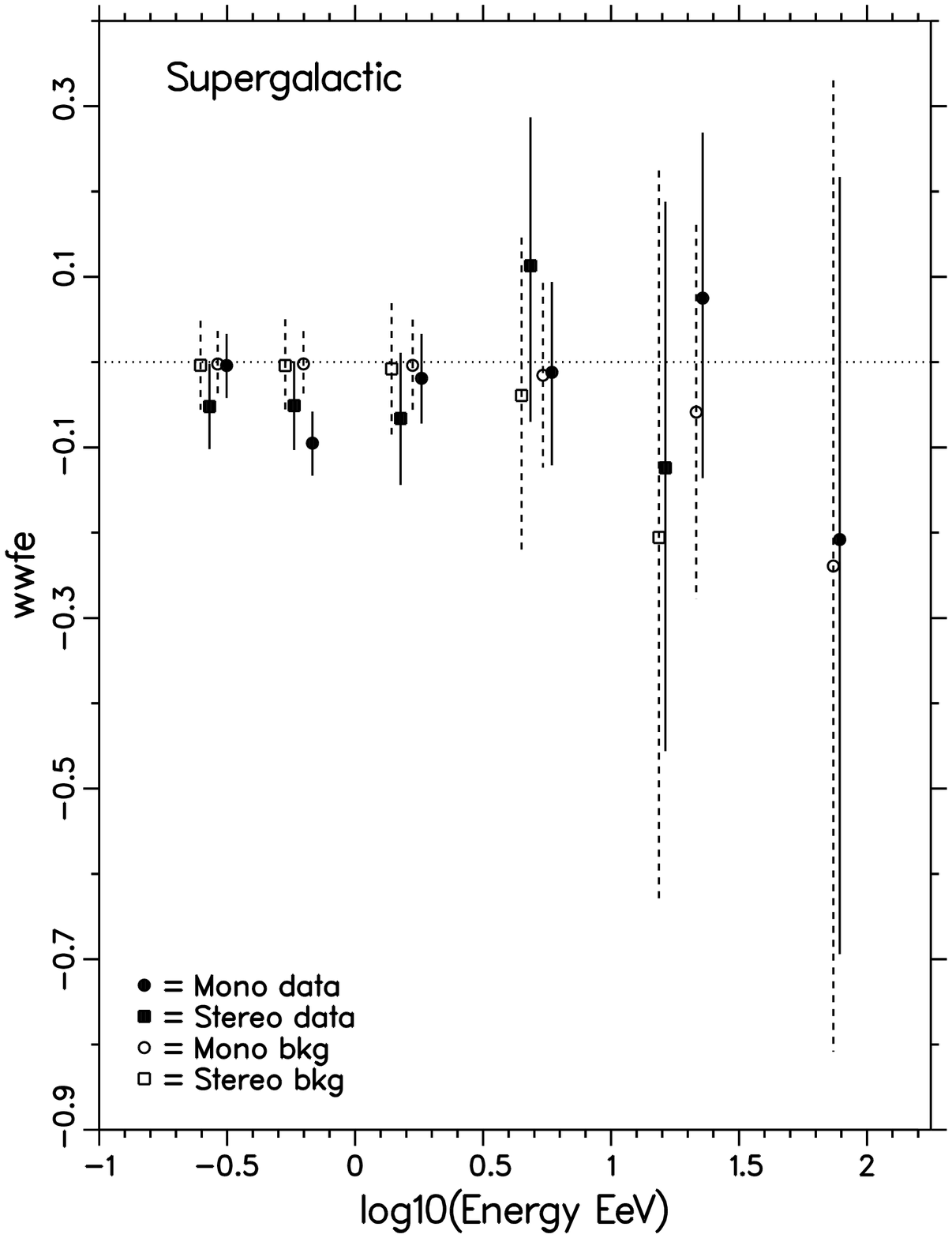]{The WWFE fit using the supergalactic latitude.}
  \label{fig:wwfe-sgb}

\figcaption[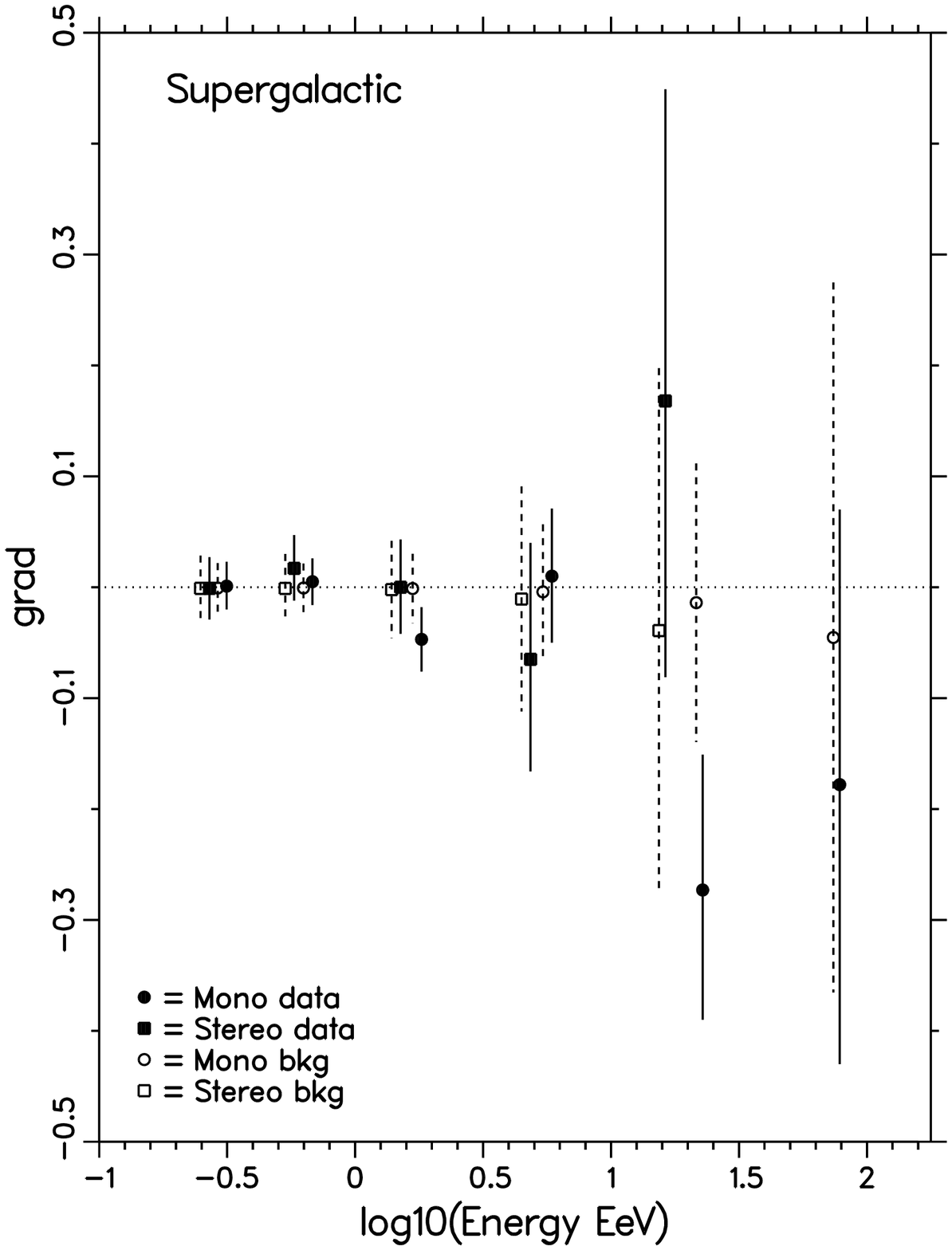]{The gradient fit using the supergalactic latitude.}
  \label{fig:grad-sgb}

\figcaption[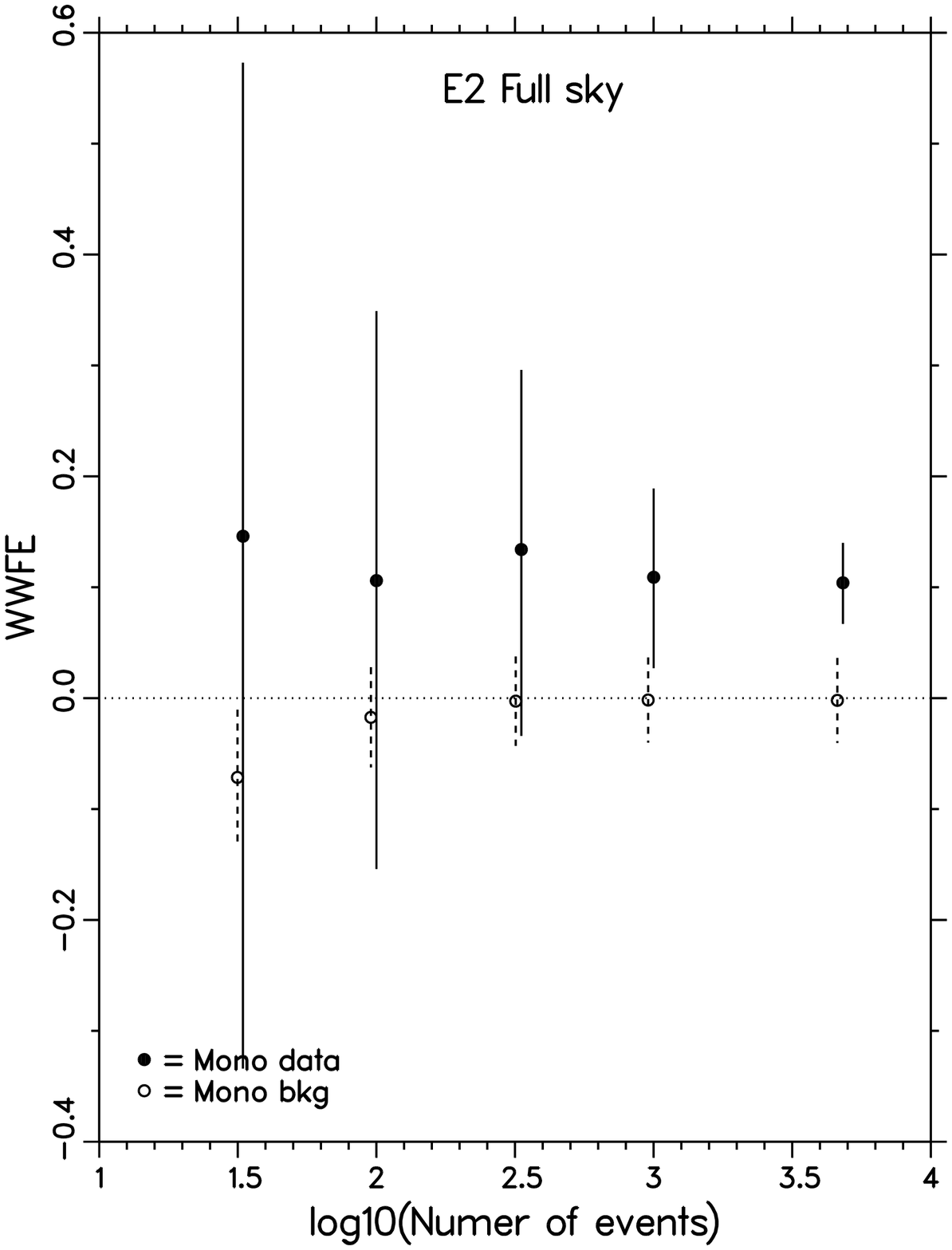]{The WWFE fit of E2 monocular data. The total number 
	of events are reduced from 4815 to 1000, 333, 100, and 33. The solid 
	dots represent the data and mean value of the simulated background are 
	given by cicles.}
  \label{fig:chk_bkg}

\figcaption[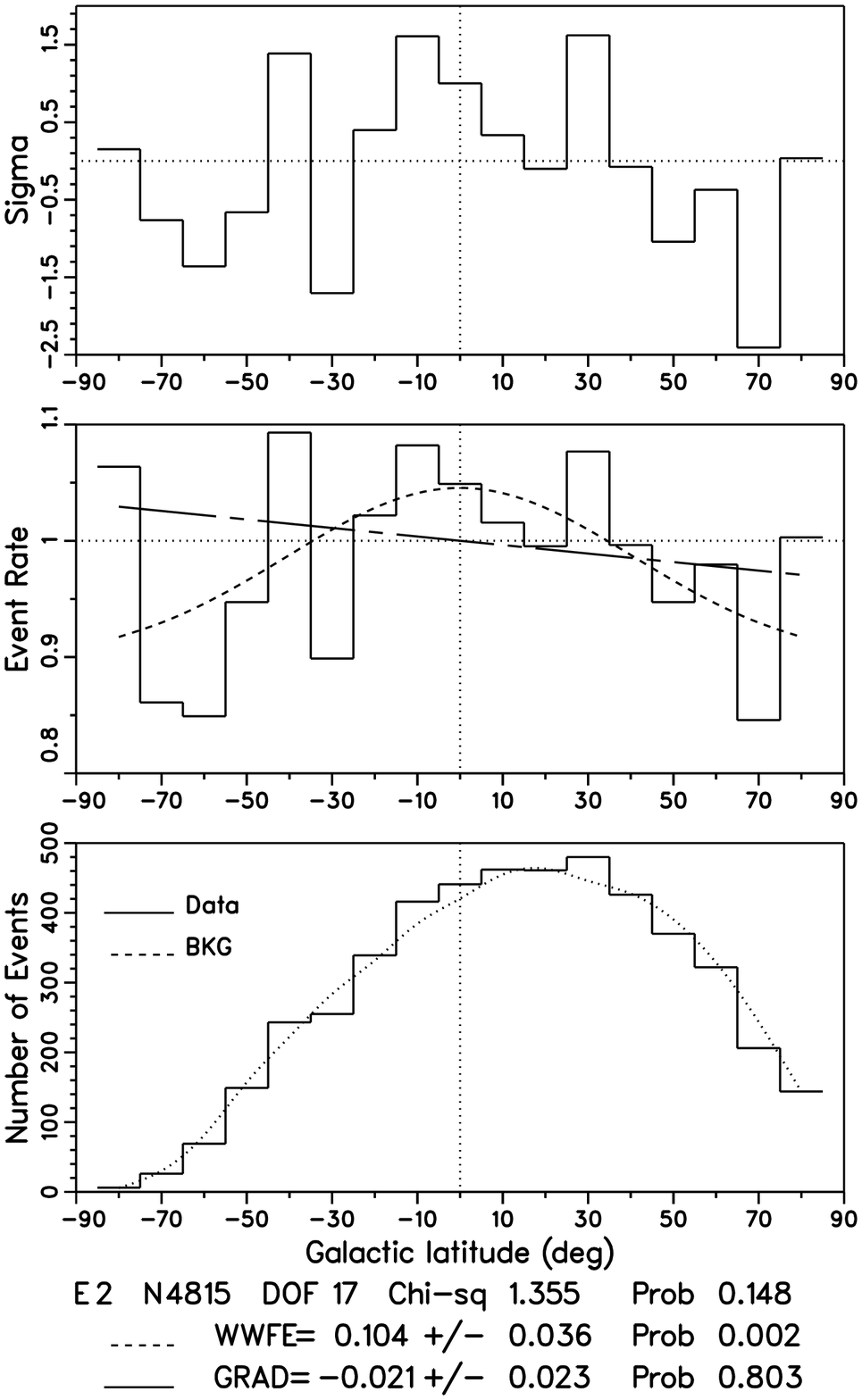]{Monocular data and isotropic background distribution 
	of the all sky zone for 0.4-1.0EeV. The bottom figure shows the
	histogram of the number of events as function of galactic latitude. The
	data ($D_i$) are represented by the histogram steps while the
	isotropic background $(B_i$) is the dotted line.  The middle figure
	show the event rate which is the ratio of number of events to
	background, $D_i/B_i$. The WWFE fit is shown by the dashed line and
	the gradient fit is shown by the dot-dashed line.The top figure show
	the significance $(D_i - B_i)/S_i$.}  \label{fig:e2z1nrs}

%\clearpage

%\plotone{match_dang.ps}

%\clearpage

%\plotone{diff_sd_z1.ps}

%\clearpage

%\plotone{e2prob.ps}

%\clearpage

%\plotone{all_wwfe.ps}

%\clearpage

%\plotone{half_wwfe.ps}

%\clearpage

%\plotone{all_grad.ps}

%\clearpage

%\plotone{half_grad.ps}

%\clearpage

%\plotone{sgb_wwfe.ps}

%\clearpage

%\plotone{sgb_grad.ps}

%\clearpage

%\plotone{chk_bkg.ps}

%\clearpage

%\plotone{e2z1nrs.ps}

\end{document}